\documentclass[prl,balancelastpage,twocolumn,footinbib,superscriptaddress,showpacs]{revtex4-1}
\usepackage{color,amsthm,amsmath,amsfonts,graphicx,bm}
\usepackage{bm}
\usepackage{color}
\usepackage{amsfonts}
\usepackage{amssymb}
\usepackage{amsmath}
\usepackage{epsfig}
\usepackage{graphicx}
\usepackage{enumerate}
\usepackage{multirow}
\usepackage{verbatim}
\usepackage{color}
\usepackage{subfigure}

\begin{document}

\title{Dynamical symmetry in quantum dissipative models}

\author{Yi Zheng}
\affiliation{State Key Laboratory of Low-Dimensional Quantum Physics and Department of Physics, Tsinghua University, Beijing 100084, China}

\author{Shuo Yang}
\email{shuoyang@tsinghua.edu.cn}
\affiliation{State Key Laboratory of Low-Dimensional Quantum Physics and Department of Physics, Tsinghua University, Beijing 100084, China}
\affiliation{Frontier Science Center for Quantum Information, Beijing, China}
\pacs{}

\begin{abstract}
 We show that the dynamical symmetry exists in dissipative quantum many-body systems. Under constraints on both Hamiltonian and dissipation parts, the time evolution of particular observables can be symmetric between repulsive and attractive interactions in the Hubbard model, or symmetric between ferromagnetic and anti-ferromagnetic interactions in the Ising model with external fields. We present a theorem to determine the existence of the dynamical symmetry in dissipative systems. This theorem is also responsible for the symmetry of steady states, even without the constraint on the initial state. We demonstrate the applications of our theorem with numerical simulations using tensor network algorithms.
 \end{abstract}

\maketitle

Dissipation occurs in systems coupled to an environment accompanied by information exchanges. As a ubiquitous effect in nature, dissipation has received great attention in quantum systems. Dissipative quantum models are at the heart of studying various novel phenomena, such as parity-time reversal symmetry breaking \cite{el2018non, li2019observation}, non-Hermitian skin effect \cite{yao2018edge,song2019non, longhi2019probing}, dissipative binding mechanisms \cite{ates2012dissipative, lemeshko2013dissipative} and phase transitions of dynamical non-equilibrium steady states \cite{carmichael1980analytical, kasprzak2006bose, amo2009collective, werner2005phase, capriotti2005dissipation, hartmann2010polariton, baumann2010dicke, nagy2010dicke, diehl2010dynamical, tomadin2011nonequilibrium, lee2011antiferromagnetic, kessler2012dissipative, hoening2012critical, lee2013unconventional, dalla2013keldysh, malossi2014full, marcuzzi2014universal, lang2015exploring, jin2016cluster, rodriguez2017probing, jin2018phase, hwang2018dissipative}. The dissipation  has also been a crucial issue in quantum engineering since it causes decoherence of quantum state. On the other hand, it can be manipulated in the preparation of particular quantum states \cite{diehl2008quantum, muller2012engineered, torosov2013non, daley2014quantum, otterbach2014dissipative}. Understanding the non-equilibrium dynamics is of common interest in a wide range of physical contexts, including ultracold gases \cite{diehl2008quantum, muller2012engineered}, trapped ions \cite{barreiro2011open, blatt2012quantum, bohnet2016quantum}, exciton-polariton Bose-Einstein condensates \cite{deng2010exciton} and cavity QED arrays \cite{tomadin2010signatures, liew2013multimode, fitzpatrick2017observation}.  

In closed quantum systems, there is an intriguing example of dynamical symmetry in Hubbard model \cite{yu2017symmetry}. The time-evolution of an observable can be symmetric between the repulsive and attractive interactions. This result is constrained by certain symmetries of the single-particle Hamiltonian. Meanwhile, the Ising Hamiltonian, as a fundamental basis for spin models, shares the same feature, \textit{i.e.}, the dynamics can be symmetric between the ferromagnetic and anti-ferromagnetic interactions. For a quantum system with dissipation, the equation of motion includes coherent parts as well as quantum jumps. A natural question arises: To what extent are the fundamental rules of symmetry in coherent dynamics maintained? Here we address this question by presenting an extended theorem that is applicable in dissipative quantum systems. The dissipative Ising and Hubbard models are demonstrated with numerical simulations by means of tensor network (TN) algorithms.

The dynamics of a dissipative quantum system is commonly governed by the master equation. This description is established by integrating over the degree of freedom of the environment \cite{pearle2012simple, brasil2013simple}. For Markovian systems, the Lindblad master equation constitutes the most general form, which is given by
\begin{equation}\label{LME}
	\frac{d}{dt} \hat\rho = -i[\hat H,\hat\rho] + \sum_\mu \left( \hat L_\mu\hat\rho \hat L_\mu^\dag-\frac{1}{2} \{\hat L_\mu^\dag \hat L_\mu,\hat\rho\} \right).
\end{equation}
Here $\hat\rho$ is the reduced density matrix of the system, $\hat H$ is the Hamiltonian, and $\hat L_\mu$ is the Lindblad operator representing the dissipation. $[\cdots]$ and $\{\cdots\}$ are respectively commutator and anti-commutator. This equation is usually written as $d\hat\rho/dt=\mathcal{\hat L}[\hat\rho]$ with $\mathcal{\hat L}$ the Liouvillian superoperator. The first term on the r.h.s. of Eq. (\ref{LME}) describes the unitary evolution as in the case of Liouville-von Neumann equation. The second term is the Lindbladian describing the non-unitary quantum jumps due to coupling to the environment. 

In general, we consider the Hermitian Hamiltonian $\hat H = \hat H_0 + \xi \hat H_\text{int}$. For the transverse-field Ising model, $\hat H_0=\sum_i\hat\sigma_i^x$ indicates linear terms of the transverse fields and $\hat H_\text{int} = \sum_{\langle i, j\rangle}\hat \sigma_i^z \hat \sigma_j^z$ is the nearest-neighbor spin interaction, with $\sigma_i^\alpha$ the $\alpha$-Pauli matrix at site $i$. Here $\xi>0$ ($\xi<0$) corresponds to the anti-ferromagnetic (ferromagnetic) interaction. For the Hubbard model, $\hat H_0$ includes single-particle hopping terms. The on-site interaction is written either as $\hat H_\text{int}=\sum_i\hat n_i(\hat n_i-1)$ for bosons or as $\hat H_\text{int}=\sum_i\hat n_{i, \uparrow}\hat n_{i, \downarrow}$ for fermions. Then $\xi>0$ ($\xi<0$) indicates the repulsive (attractive) interaction. The dynamical symmetry here refers to the phenomenon that a certain observable $\mathcal{\hat O}$ shares the symmetric time-evolution for models with $+\xi$ and $-\xi$, \textit{i.e.}, $\langle\mathcal{\hat O}(t)\rangle_{+\xi}=\pm\langle\mathcal{\hat O}(t)\rangle_{-\xi}$. 

In unitary evolutions with $\hat L_\mu = 0$, the dynamical symmetry of the Hubbard model has been proven to be reflected by the single-particle Hamiltonian \cite{yu2017symmetry}. As a pure quantum state evolves as $|\psi(t)\rangle = e^{-i(\hat H_0+\xi\hat H_\text{int})t}|\psi_0\rangle$, the combination of time reversal and a symmetry operation (that flips the sign of $\hat H_0$) leads to the dynamics of an effective Hamiltonian $\hat H_0-\xi\hat H_\text{int}$. We take the transverse-field Ising model $\hat H = (h_x/2)\sum_i\hat \sigma_i^x+(J/4)\sum_{\langle i, j\rangle}\hat \sigma_i^z\hat \sigma_j^z$ as an example. The physics is unchanged under a $\pi$-rotation around the $z$-axis in spin space ($\hat W^{-1}\hat\sigma_x\hat W=-\hat\sigma_x$ and $\hat W^{-1}\hat\sigma_y\hat W =-\hat\sigma_y$ with $\hat W=e^{\pm\frac{i}{2}\pi\hat\sigma_z}$). By applying time-reversal followed by this rotation, we can map the evolution operator $e^{-i\hat Ht}$ of an anti-ferromagnetic model to that of a ferromagnetic model. Thus as long as the initial state satisfies $(\hat R\hat W)^{-1}|\psi_0\rangle=e^{i\phi}|\psi_0\rangle$ with $\hat R$ being the time-reversal operator ($\hat R^{-1}i\hat R=-i$) and $\phi$ an arbitrary phase, we have $\langle\mathcal{\hat O}(t)\rangle_{+J}=\langle\mathcal{\hat O}(t)\rangle_{-J}$ for $\mathcal{\hat O}=\hat \sigma_z,\hat \sigma_y$ and $\langle\mathcal{\hat O}(t)\rangle_{+J}=-\langle\mathcal{\hat O}(t)\rangle_{-J}$  for $\mathcal{\hat O}=\hat \sigma_x$, with ``$+$" or ``$-$" in the subscripts indicating the sign of $J$. 

In contrast to closed quantum systems, the dynamical symmetry for a dissipative model requires restrictions on the jump operator $\hat L_\mu$ and the initial density matrix $\hat\rho_0$. We present our theorem here and provide the proof subsequently. The appearance of dynamical symmetry depends on the existence of an anti-unitary operator $\hat S$ that satisfies the following conditions: (i) $\{\hat S, \hat H_0\} = 0$ and $[\hat S,\hat H_\text{int}] = 0$, (ii) $[\hat S, \hat \rho_0] = 0$, (iii) $\hat S^{-1}\mathcal{\hat O} \hat S = \pm\mathcal{\hat O}$ with $\mathcal{\hat O}$ the observable operator and (iv) $\hat S^{-1}\hat L_\mu \hat S = e^{i\phi}\hat L_\mu$ with $\phi$ an arbitrary phase. Here conditions (i-iii) are responsible for the dynamical symmetry in unitary evolutions as demonstrated in Ref. [\onlinecite{yu2017symmetry}]. Condition (iv) indicates the restriction on the dissipative parts. Upon finding such an operator $\hat S$, we can conclude that $\langle\mathcal{\hat O}(t)\rangle_{+\xi}=\pm\langle\mathcal{\hat O}(t)\rangle_{-\xi}$, with the ``$\pm$" sign in accordance with that in (iii). In practice, it is convenient to express $\hat S = \hat R\hat W$ with $\hat R$ the anti-unitary time-reversal operator and $\hat W$ a unitary operator.

To proof this theorem, we notice that Eq. (\ref{LME}) yields a formal solution $\hat\rho(t) = e^{\mathcal{\hat L}t}\hat\rho_0$. We obtain 
\begin{eqnarray}\label{Obs}
	\langle\mathcal{\hat O}(t)\rangle&=&\text{Tr}\left(e^{\mathcal{\hat L}t}\hat\rho_0\mathcal{\hat O}\right)
	=\text{Tr}\left(\hat S^{-1}e^{\mathcal{\hat L}t}\hat S\hat S^{-1}\hat \rho_0\hat S\hat S^{-1}\mathcal{\hat O}\hat S\right)\nonumber\\
	&=&\pm\text{Tr}\left(e^{\hat S^{-1}\mathcal{\hat L}\hat St}\hat\rho_0\mathcal{\hat O}\right)
\end{eqnarray}
from conditions (ii) and (iii). Then we need to prove $\hat S^{-1}\mathcal{\hat L}(+\xi)\hat S=\mathcal{\hat L}(-\xi)$ with conditions (i) and (iv). We consider the so-called ``Choi's isomorphism", which can be intuitively understood by the mapping $|\psi\rangle\langle\phi|\leftrightarrow|\psi\rangle\otimes|\phi\rangle$ \cite{zwolak2004mixed, kshetrimayum2017simple}. The insight is to vectorize (reshape) the density matrix $\hat\rho$ into a super-ket state $|\rho\rangle_\#$ and rewrite Eq. (\ref{LME}) as $|\dot\rho\rangle_\# = \mathcal{\hat L}_\#|\rho\rangle_\#$. Here 
\begin{eqnarray}\label{Lind}
	\mathcal{\hat L}_\#&=&-i\left(\hat H\otimes \hat I- \hat I\otimes \hat H^T\right)\nonumber\\
	&+&\sum_\mu\left( \hat L_\mu\otimes \hat L_\mu^*-\frac{1}{2} \hat L_\mu^\dag \hat L_\mu\otimes \hat I-\frac{1}{2} \hat I\otimes \hat L_\mu^T \hat L_\mu^*\right)
\end{eqnarray}
is the ``vectorized" Liouvillian superoperator \cite{kshetrimayum2017simple}. Operators on the two sides of ``$\otimes$" respectively act on the ket and the bra of $\hat \rho$. $\hat I$ is a $d\times d$ identity matrix with $d=\dim(\hat H)$. In addition, we have the mapping
\begin{eqnarray}
	\text{Tr}(\hat \rho\mathcal{\hat O})\leftrightarrow _\#\langle\mathcal{O}|\rho\rangle_\#,\ 
	\hat S^{-1}\mathcal{\hat O}\hat S\leftrightarrow \mathbb{\hat S}^{-1}|\mathcal{O}\rangle_\#,
\end{eqnarray}
with $|\mathcal{O}\rangle_\#$ the vectorized observation operator and $\mathbb{\hat S}^{-1} \equiv (\hat S^{-1}\otimes \hat S^T)$. Now Eq. (\ref{Obs}) can be rewritten as
\begin{eqnarray}\label{Proof}
	\langle\mathcal{\hat O}(t)\rangle &=& _\#\langle\mathcal{O}|e^{\mathcal{\hat L}_\#t}|\rho_0\rangle_\#
	= _\#\langle\mathcal{O}|\mathbb{\hat S}\mathbb{\hat S}^{-1}e^{\mathcal{\hat L}_\#t}\mathbb{\hat S}\mathbb{\hat S}^{-1}|\rho_0\rangle_\#\nonumber\\
	&=& \pm_\#\langle\mathcal{O}|e^{\mathbb{\hat S}^{-1}\mathcal{\hat L}_\#\mathbb{\hat S}\;t}|\rho_0\rangle_\#.
\end{eqnarray}
From Eq. (\ref{Lind}), we can straightforwardly verify that $\mathbb{\hat S}^{-1}\mathcal{\hat L}_\#(+\xi)\mathbb{\hat S}=\mathcal{\hat L}_\#(-\xi)$ under conditions (i) and (iv). This leads to $\langle\mathcal{\hat O}(t)\rangle_{+\xi}=\pm\langle\mathcal{\hat O}(t)\rangle_{-\xi}$ from Eq. (\ref{Proof}).

Furthermore, the dissipative dynamics of Eq. (\ref{LME}) commonly leads to a non-equilibrium steady state characterized by $\frac{d}{dt}\hat\rho_{ss}=0$, with $\hat\rho_{ss}=\lim_{t\to\infty}\hat\rho(t)$. Consider that the stationary state is unique, which is indeed the case under quite general assumptions \cite{albert2014symmetries, minganti2018spectral, nigro2019uniqueness}. Upon finding an anti-unitary operator $S$ that satisfies conditions (i) and (iv), we have $\mathcal{\hat L}[\hat \rho_{ss}]_{+\xi}=\mathcal{\hat L}[\hat S^{-1}\hat \rho_{ss}\hat S]_{-\xi}=0$ with $\hat\rho_{ss}$ the steady state corresponding to $\mathcal{\hat L}(+\xi)$. Thus $\hat S^{-1}\hat \rho_{ss}\hat S$ is the steady state for $\mathcal{\hat L}(-\xi)$. If we further have condition (iii) $\hat S^{-1}\mathcal{\hat O} S = \pm\mathcal{\hat O}$, we can conclude that 
\begin{equation}\label{SS}
	\langle\mathcal{\hat O}\rangle_{ss,+\xi}=\text{Tr}(\hat S^{-1}\hat \rho_{ss}\hat S\hat S^{-1}\mathcal{\hat O}\hat S) = \pm\langle\mathcal{\hat O}\rangle_{ss,-\xi}, 
\end{equation}
with $\langle\mathcal{\hat O}\rangle_{ss,\pm\xi}$ the expectation value of $\mathcal{\hat O}$ in steady state for $\mathcal{\hat L}(\pm\xi)$. Eq. (\ref{SS}) is established without the constraint on the initial density matrix, since a steady state (if existed) should be independent of the initial state of the dynamics. This conclusion can be quite helpful in analysing phases of non-equilibrium steady states, which is in the similar spirit of the symmetry analysis for closed systems.

On the other hand, there have been numerical efforts in simulating dissipative quantum many-body systems \cite{weimer2015variational, cui2015variational, werner2016positive, gangat2017steady}. To perform time-evolutions, we take the recent proposed TN method which is based on the TN representation of a quantum state/operator \cite{kshetrimayum2017simple}. The main idea is to map an operator (such as Projected Entangled-Pair Operator, or PEPO) to a TN state (such as Projected Entangled-Pair State, or PEPS \cite{verstraete2008matrix,orus2014practical}) according to the ``Choi's isomorphism". This is done by binding the two physical indices into a single one through $(i,j)\to j\cdot d+i$, with $i$ and $j$ the row and column indices of the operator in its matrix form. Thus the traditional TN algorithms can be applied. The infinite-PEPS (iPEPS) algorithm has been demonstrated to be reliable in simulating the steady state of the two-dimensional (2D) square lattice systems in the thermodynamic limit \cite{kshetrimayum2017simple}. Here we use the Projected Entangled-Simplex Operator (PESO) TN \cite{xie2014tensor} as a representation of the density matrix and apply the simple update algorithm \cite{vidal2007classical,jiang2008accurate}. For infinite 2D systems, the TN is translationally-invariant and we take a $2\times2$ unit cell. This method is efficient and stable in real-time simulations with bond dimensions $D=6$ for the spin-1/2 model and $D=4$ for the Bose-Hubbard model. A time step $\delta t=0.1\sim0.01$ is sufficient in our simulations. We apply the Corner Transfer Matrix method (CTM) \cite{orus2009simulation} to contract the tensor network for the expectation value of an observable. Convergence of the results have been verified with the truncated bond dimension $\chi=10$ in CTM.

Now we consider a dissipative Ising model on an infinite 2D square lattice as a demonstration of our theorem. We focus on the spin-1/2 quantum system with incoherent dissipation given by $\hat L_\mu = \sqrt \gamma\hat\sigma_\mu^-$ that flips one spin. Here $\gamma=0.1$ denotes the decay rate of the up-spins and $\hat\sigma_\mu^-=\hat\sigma_\mu^x-i\hat\sigma_\mu^y$ is the spin lowering operator with $\mu$ the site index. The dynamics is governed by the master equation (\ref{LME}) with the Hamiltonian given by 
\begin{equation}\label{Hamil}
	\hat H = \frac{h_x}{2}\sum_i\hat \sigma_i^x + \frac{h_z}{2}\sum_i\hat \sigma_i^z + \frac{J}{4}\sum_{\langle i, j\rangle}\hat \sigma_i^z\hat \sigma_j^z.
\end{equation}
Here $J$ is the interaction strength and $h_x$ ($h_z$) is the strength of the transverse (longitudinal) field. This model has received theoretical and experimental interests due to its connection to the systems of driven-dissipative Rydberg gases \cite{lee2011antiferromagnetic, carr2013nonequilibrium, weimer2015variationalana, maghrebi2016nonequilibrium, overbeck2017multicritical}. In previous studies, calculations of steady state have exhibited a first order phase transition akin to the liquid-gas transition \cite{weimer2015variational, weimer2015variationalana, overbeck2017multicritical} (despite early doubts on the existence of a bistable phase \cite{lee2011antiferromagnetic, marcuzzi2014universal}). The up-spins are treated as particles and its density $n_\uparrow=\sum_{i=1}^N\langle(\hat 1+\hat \sigma_i^z)\rangle/2N$ (with $i$ runs over all lattice sites) is the order parameter. 

\begin{figure}[tbp]
  \includegraphics[width=8.5cm, height=8.5cm]{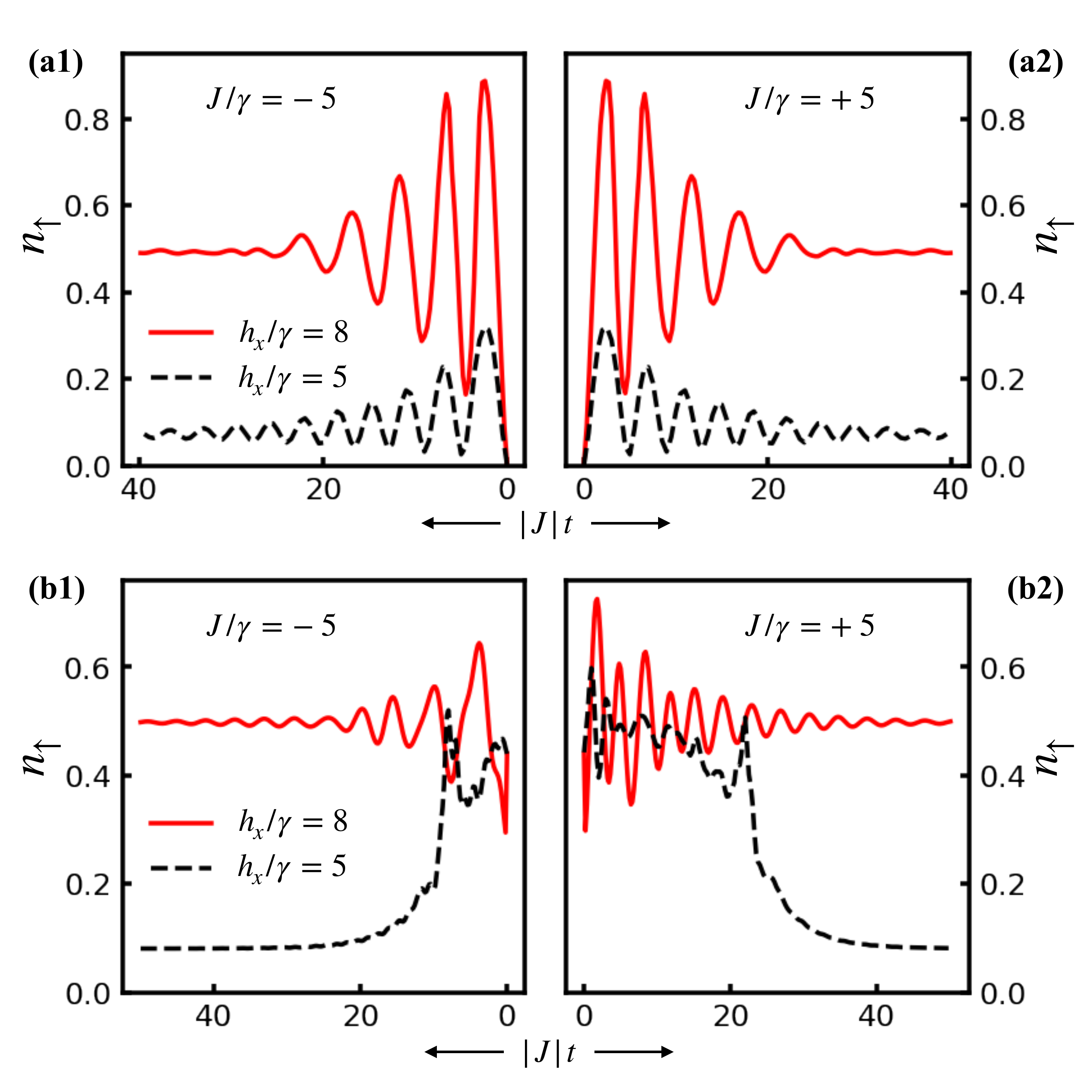}
  \caption{Time evolution of spin-up density $n_\uparrow$ in the dissipative transverse-field Ising model. (a1, a2) Symmetric evolutions with ferromagnetic interaction ($J/\gamma = -5$) and anti-ferromagnetic interaction ($J/\gamma = 5$). The upper and lower curves correspond to $h_x/\gamma = 8$ and $5$, respectively. The evolutions start from a polarized state with all spins pointing down. (b1, b2) The same as (a1, a2) but with the initial set to be a random mixed state. The dynamical symmetry is not satisfied. However, in long-time evolutions, spin-up densities in steady states remain symmetric between $J/\gamma = -5$ and $J/\gamma = +5$. }
  \label{fig1}
\end{figure}

\begin{figure}[tbp]
  \includegraphics[width=8.5cm, height=8cm]{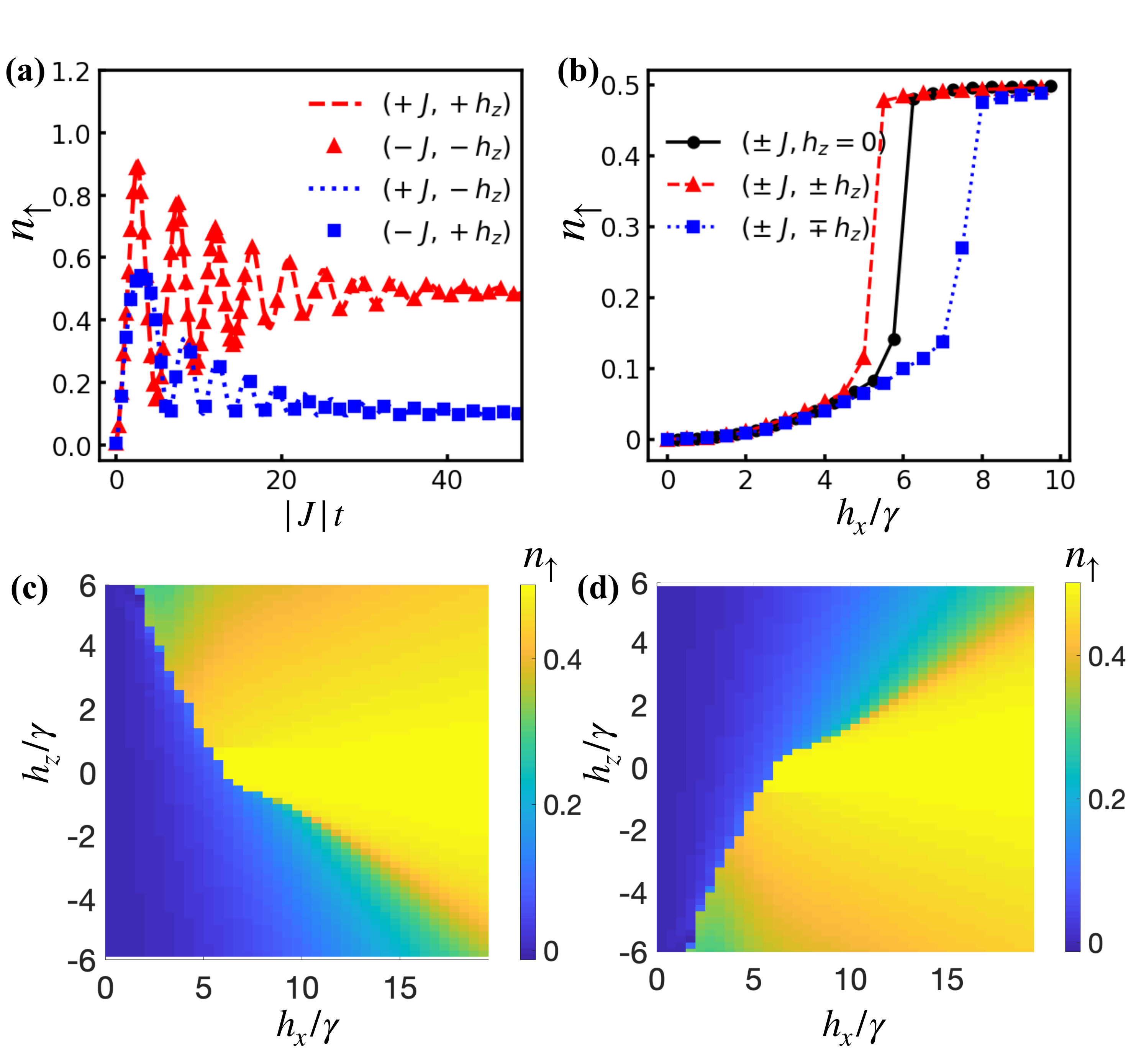}
  \caption{(a) Time dependence of spin-up density $n_\uparrow$ in the dissipative Ising model with transverse field ($h_x/\gamma = 6.5$) and longitudinal field. $n_\uparrow(t)$ obtained with parameters ($+J$,$+h_z$) (red dashed line) differs from that for ($-J$,$+h_z$) (blue square), but it agrees with the data of ($-J$,$-h_z$) (red triangle). (b) Spin-up density $n_\uparrow$ in the steady state as a function of $h_x$. Steady states for $(+J,\pm h_z)$ and $(-J,\mp h_z)$ share the same $n_\uparrow$. (c, d) The steady state phase diagram spanned by $h_x$ and $h_z$. The two diagrams in (c) and (d) are for anti-ferromagnetic ($+J$) and ferromagnetic ($-J$) models. They are symmetric about $h_z = 0$. We take $\pm J = \pm 5\gamma$ and $\pm h_z = \pm 0.5\gamma$ in all plots.}
  \label{fig2}
\end{figure}

We first consider the case of $h_z=0$ and hence $\hat H_0 = (h_x/2)\sum_i\hat \sigma_i^x$. One can find $\hat S=\hat R(\otimes_i\hat \sigma_i^z)$ such that $[\hat S,\hat\sigma_i^z\hat\sigma_j^z]=0$, $\{\hat S, \hat\sigma_i^x\}=0$, $[\hat S,(\hat 1+\hat \sigma_i^z)]=0$ and $\{\hat S, \hat\sigma^-\}=0$ in accordance with conditions (i), (iii) and (iv). In general, condition (ii) is satisfied if we assume the initial state to be of the form $\hat \rho_0=\sum_nP_n|\phi_n\rangle\langle\phi_n|$, where $|\phi_n\rangle = |s_1,\cdots,s_N\rangle_n$ with $|s_i\rangle$ an eigenstate of $\hat \sigma^z$.  We immediately conclude that the time-evolutions of $n_\uparrow$ are symmetric between the anti-ferromagnetic ($J>0$) and ferromagnetic  ($J<0$) interactions. As an example, we choose $\hat \rho_0=|\downarrow\cdots\downarrow\rangle\langle\downarrow\cdots\downarrow|$. As can be seen from Fig.~\ref{fig1} (a1) and (a2) with $J = -5\gamma$ and $J = 5\gamma$, the spin-up densities $n_\uparrow$ follow exactly the same evolution. For $h_x = 5\gamma$ (the lower curves), the dynamics leads to a steady state with low density of up-spins, which corresponds to the lattice gas phase. The steady state for $h_x = 8\gamma$ (the upper curves) indicates a high-density liquid phase, characterized by nearly half-filling up-spins and vanishing compressibility ($-\partial n_\uparrow/\partial h_x$) \cite{weimer2015variational}. The phase transition is shown in Fig.~\ref{fig2} (b). Note that the initial polarized state $\hat\rho_0$ is also the steady state for $h_x=0$, since the dissipation term will eventually flip every spin down. On the other hand, if we start the evolution from another steady state obtained with finite $h_x\neq 0$, the dynamical symmetry is broken. Even if condition (ii) is violated, we can still conclude from Eq. (\ref{SS}) that the spin-up densities in steady states are symmetric between $J>0$ and $J<0$. As a demonstration, we set a random density matrix $\hat\rho_r$ as the initial state and perform the simulations with $J = \pm5\gamma$, $h_x = 5\gamma$ and $8\gamma$. Results are shown in Fig.~\ref{fig1} (b1) and (b2). Obviously, $\hat\rho_0=\hat\rho_r$ does not commute with $\hat S$ and hence the dynamical symmetry is broken. In long-time evolutions, however, the densities of up-spins lead to the same value for both $J=5\gamma$ and $-5\gamma$, as expected.

In the presence of longitudinal field $(h_z/2)\sum\hat \sigma_i^z$, we include this term into the spin interaction part $\hat H_\text{int}$ instead of $\hat H_0 = (h_x/2)\sum\hat \sigma_i^z$, our theorem still applies with the same anti-unitary operator $\hat S$ as in the case of $h_z=0$. Now the conclusion is about the symmetry between $(+J,\pm h_z)$ and $(-J, \mp h_z)$, $\textit{i.e.}$, between  anti-ferromagnetic interaction with positive (negative) longitudinal field and ferromagnetic interaction with negative (positive) longitudinal field. As shown in Fig.~\ref{fig2} (a) with $h_x = 6.5\gamma$, $\pm h_z = \pm0.5\gamma$ and $\pm J = \pm 5\gamma$, although the time dependence of $n_\uparrow$ for $(+J,+h_z)$ differs from that for $(-J,h_z)$, it is equivalent to the case of $(-J,-h_z)$. As the dynamical symmetry sufficiently leads to the symmetry of stable steady states, such symmetric behavior also appears in the steady state phase diagram. In Fig.~\ref{fig2} (b), the spin-up densities $n_\uparrow$ (in steady states) as functions of $h_x$ indicates the first-order transition from lattice gas to lattice liquid \cite{weimer2015variational,jin2018phase}. For $J=5\gamma$ and $h_z=0$, the solid curve with the transition point at $h_x\sim6\gamma$ is in good agreement with the results from both the variational method with correlated ansatz \cite{weimer2015variational} and the real-time simple update with iPEPS \cite{kshetrimayum2017simple}. The transition point shifts leftwards (rightwards) when $h_z=0.5\gamma$ ($-0.5\gamma$). We observe that the $n_\uparrow$-$h_x$ relations with $(J,h_z)/\gamma=(5,0)$, $(5,0.5)$ and $(5,-0.5)$ are reproductions of those with $(J,h_z)/\gamma=(-5,0)$, $(-5,-0.5)$ and $(-5,0.5)$, respectively. We complement our demonstration with the phase diagram spanned by $h_x$ and $h_z$, as depicted in Fig.~\ref{fig2} (c) for $J = 5\gamma$ and (d) for $J = -5\gamma$. One can clearly see that the two diagrams are symmetric about the ($h_z = 0$)-axis. The symmetry of phase diagrams between the anti-ferromagnetic and ferromagnetic models relies on the dissipative nature of the system, which is remarkably different from the situation in closed quantum systems. For example, in non-dissipative Ising models with a small transverse field, the anti-ferromagnetic and ferromagnetic interactions correspond to distinct phases that are  characterized by $n_\uparrow = 1/2$ and $1$, respectively.

To demonstrate the application of our theorem in dissipative Hubbard model, we turn to a bosonic system in optical lattice with two-body dissipation. This consideration is motivated by the recent experimental realization of such a model and its high controllability \cite{tomita2017observation}. The experiment in Ref. [\onlinecite{tomita2017observation}] implemented a slow ramp down of lattice depth to observe the crossover from Mott insulator (MI) to superfluid (SF) state. Without loss of generality, here we consider the quench dynamics in a 2D square lattice and show that the time evolutions of the average density are symmetric between the repulsive and attractive models. Note that the analysis can be directly extended to other realizable configurations including the Harper-Hofstadter Hamiltonian \cite{tai2017microscopy} and the fermionic Aubre-Andr\'e model \cite{schreiber2015observation}.  

\begin{figure}[tbp]
  \includegraphics[width=8.5cm]{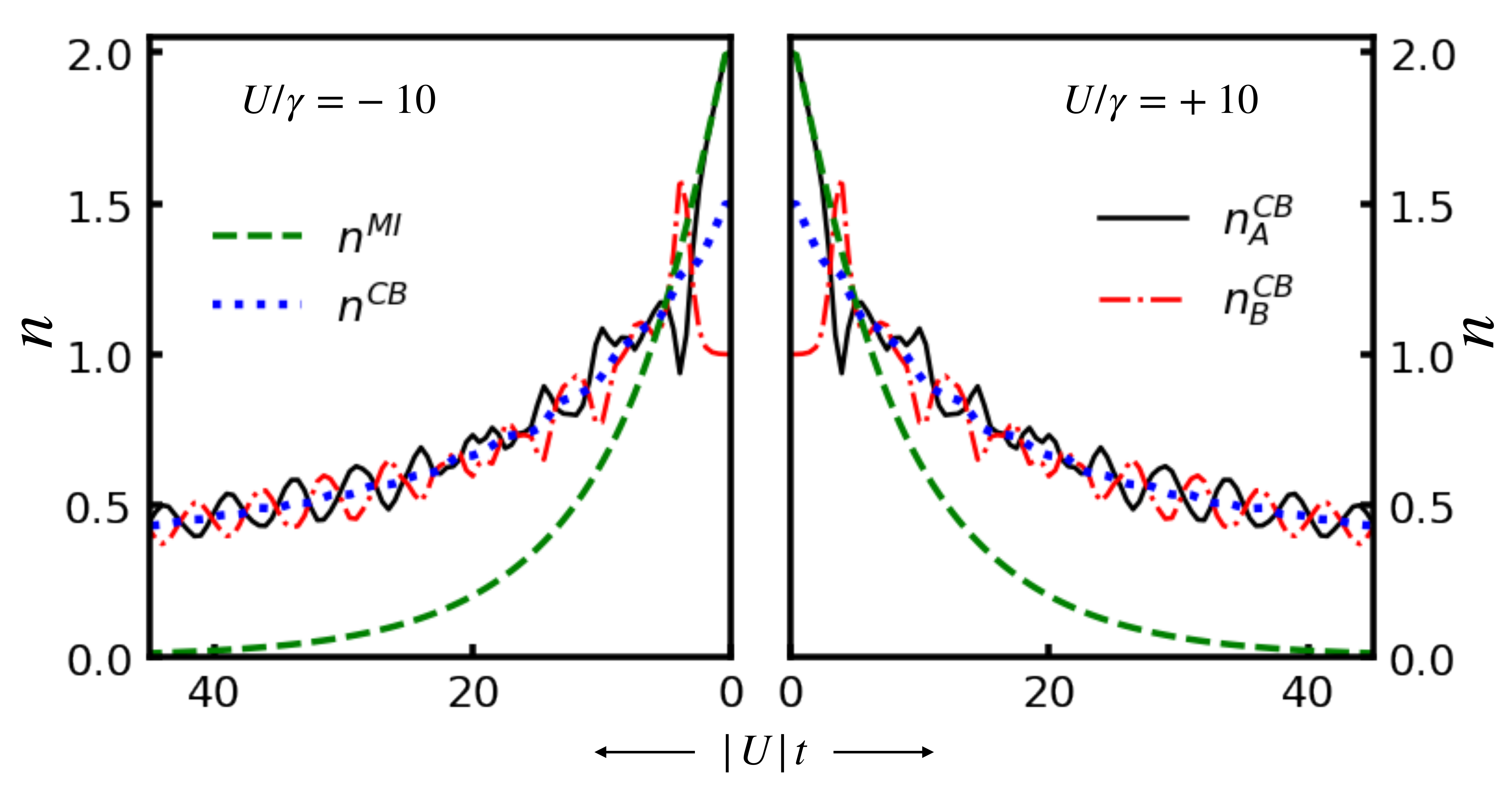}
  \caption{Decay of the average density $n$ in the Bose-Hubbard model with two-body atom loss. The evolutions start from a double-filled MI or a $(2,1)$-filled CB state. The time-dependent densities are symmetric between the attractive (left panel with $U = -5$) and repulsive (right panel with $U = +5$) models. Here $n_A^\text{CB}$ and $n_B^\text{CB}$ are neighboring-site densities starting form the CB state.}
  \label{fig3}
\end{figure}

The two-body inelastic atom loss can be engineered through a single photon photoassociation process for ultracold gases in optical lattices \cite{tomita2017observation}, which leads to the jump operator of the form $\hat L_\mu = \sqrt{\gamma}\hat b_\mu\hat b_\mu$. The Hamiltonian is given by
\begin{equation}\label{BHM}
	\hat H = -\sum_{\langle j,l\rangle}\hat b_j^\dag \hat b_l + \frac{U}{2}\sum_j \hat n_j(\hat n_j-1),
\end{equation}
with $\hat b_j$ ($\hat b_j^\dag$) the bosonic annihilation (creation) operator and $\hat n_j = \hat b_j^\dag \hat b_j$. It has been pointed out that in a closed system with Hamiltonian (\ref{BHM}), the dynamical symmetry corresponds to the unitary operator $\hat W$ given by $\hat W^{-1} \hat b_j \hat W=(-1)^j \hat b_j$. Then $\hat S = \hat R \hat W$ commutes with the interaction term and anti-commutes with $\hat H_0 = -\sum_{\langle j,l\rangle}\hat b_j^\dag \hat b_l$. Thus a bipartite lattice geometry is essential in applying this theorem. As the explicit form of $\hat W$ is not given, here we present $\hat W = \otimes_j\exp[i\frac{\pi}{2}\hat n_j+(-1)^ji\frac{\pi}{2}\hat n_j]$ that meets the requirements. As a result, the relation $\hat S^{-1}\hat L_\mu S = e^{i\phi}\hat L_\mu$ is always fulfilled for on-site dissipations consist of $\hat b_j$ ($\hat b_j^\dag$). In general, we take the initial state as $\hat \rho_0 = \sum_mP_m|\phi_m\rangle\langle\phi_m|$ with $|\phi_m\rangle=|n_1,\cdots,n_N\rangle_m$, and the physical operator as $\mathcal{\hat O}=(1/N)\sum_{j=1}^N\hat n_j$. It is straightforward to verify that $\hat S$ commutes with $\hat \rho_0$ and $\mathcal{\hat O}$. Now with conditions (i-iv) all satisfied in our consideration, the dynamical symmetry $\langle\mathcal{\hat O}(t)\rangle_{+U}=\langle\mathcal{\hat O}(t)\rangle_{-U}$ is guaranteed. The time dependence of the average density is illustrated in Fig.~\ref{fig3} with $\gamma=0.5$, $U = \pm 5$. The maximum occupation on each site is limited to 3. The evolutions start from a double-filled MI and a checkerboard (CB) state with filling factors $n_A=2$ and $n_B=1$ on neighboring sites. Both cases lead to identical density decay in attractive and repulsive interacting systems. 

In summary, we have provided a theorem to capture the insight of the dynamical symmetry in dissipative quantum many-body systems. Demonstrations with both Ising model and Bose-Hubbard model indicate that the symmetric behaviors in time evolutions are connected to the intrinsic features of the dissipation system. In contrast to closed systems, the symmetry in steady states is valid even without the constraint on the initial state. This can be taken as an advantage in analyzing the steady-state phase diagram. Our results can be further applied to experiments with Rydberg polaritons, optical cavities, as well as solid-state materials. Besides, as dissipation is ubiquitous in nature, this universal theorem will provide important guiding principles for controlling dissipation and decoherence in many quantum systems.

This work is supported by the National Key R\&D Program of China (Grant No. 2018YFA0306504) and NSFC (Grant No. 11804181). S. Y. acknowledges supports from the National Thousand Young Talents Program and Tsinghua University Initiative Scientific Research Program.
\bibliography{DS}

\end{document}